

\input phyzzx.tex

\def\quadg {g^{ab}\nabla_a\nabla_b}
\def\quadgg {\hat g^{ab}\nabla_a\nabla_b}

\def\qgg {\Delta_{\hat g}}
\def\d {\partial}
\def\dd {\bar\partial}
\def\y {\hat g}
\def\v {\varphi}
\def\j {\lambda}
\def\a {\alpha}
\def\g {g_{ab}}
\def\h {\hat g_{ab}}

\titlepage
Dipartimento di Fisica\par
Universit\`a di Roma "La Sapienza"\par
I.N.F.N. -- Sezione di Roma \par\par
{\bf Preprint n.853} \par
hepth@xxx/9201069 \par

\title{A continuum string model for $D>1$}

\author {M. Martellini\foot{On leave of absence from Dipartimento di Fisica,
Universit\'a
di Milano, Milano, Italy and I.N.F.N., Sezione di Pavia, Italy}}
\address{I.N.F.N., Sezione di Roma "La Sapienza", Roma, Italy}
\author {M. Spreafico}
\address{Dipartimento di Fisica, Universit\'a di Milano, Milano, Italy}
\author {K. Yoshida}
\address{Dipartimento di Fisica, Universit\'a di Roma "La Sapienza", Roma,
Italy and I.N.F.N., Sezione di Roma, Italy}

\abstract
The non critical string (2D gravity coupled to the matter with central
charge $D$) is quantized taking care of both diffeomorphism and Weyl
symmetries.
In incorporating the gauge fixing with respect to the Weyl symmetry, through
the condition $R_g=const$, one modifies the classical result of Distler and
Kawai. In particular one obtains the real string tension for an arbitrary value
of central charge $D$.
\endpage

\pagenumber=1

Conformal field theory coupled to 2D Euclidean quantum
gravity was solved initially by
Polyakov [1] who introduced the following path integral
$$
Z=\int{[D_gX^{\mu}][D_gg^{ab}] \over Vol(Diff)~Vol(Weyl)}
      e^{-S_M[g_{ab};X^\mu]-{\mu_{\scriptscriptstyle{0}} \over 2\pi}
\int d^2\xi\sqrt{g}}
\eqn\uno
$$
where the matter action is
$$
S_M[g;X]={1 \over 8\pi} \int d^2\xi\sqrt{g}g^{ab}\nabla_aX^\mu\nabla_bX^\mu
$$
and an explicit bare cosmological constant term has been included.
The $X^\mu$ are $D$ bosonic scalar fields taking values in $\Re^D$.

As well known the presence of the volumes in the measure follows from the
invariance of the matter action under diffeomorphisms and conformal mappings
of the Riemann surface [2] [3].
Thus we are lead to analyse the symmetries of matter action. As is known
$S_M[g;X]$ is invariant both under

\noindent
- diffeomorphism of the word sheet
$$
\xi^a\to\xi'^a=\xi^a+\epsilon^a(\xi)
$$
$$
\delta g_{ab}=\nabla_a\epsilon_b+\nabla_b\epsilon_a
$$

\noindent
- Weyl transformation, that is transformations of the metric field alone
of the type
$$
g'_{ab}(x)=e^{\sigma(x)}g_{ab}
$$

Our purpose is to fix the gauge and define the correct determinant
and Feedov Popov ghosts for both the gauge symmetries. Furthermore we shall
assume to have a string world-sheet surface of genus $h\ge 2$, i.e.
an hyperbolic 2D-geometry.

To fix Weyl gauge we note that for every given hyperbolic Euclidean
metric $g_{ab}$, there exists
only one metric $g'_{ab}$ conformal to $g_{ab}$ having constant curvature
$R_{g'}=-1$ [4] (see appendix A).
So we can choose the gauge fixing condition
$$
R_g=-1
$$

To write the FP determinant we must calculate the formal derivative of
the gauge function
$$
f(g'_{ab})=R_{g'}+1
$$
when $g'_{ab}$ is given by the Weyl transformation with a scaling
parameter $k$
$$
g'_{ab}(x)=e^{k\sigma(x)}g_{ab}
$$
so
$$
f(g')=e^{-k\sigma}(R_g - kg^{ab}\nabla_a\nabla_b\sigma)+1
$$

If we consider an infinitesimal variation $\delta\sigma$, we have
$$
\eqalign {&\delta g_{ab}=kg_{ab}\delta\sigma\cr
& R_{g'}=(1-k\delta\sigma)(R_g - k\quadg\delta\sigma)=
R_g - k\quadg\delta\sigma - kR_g\delta\sigma + o(\delta\sigma^2)\cr
& \delta R_g = -k(R_g + \quadg)\delta \sigma\cr
& {\delta f \over \delta\sigma} = -k(R_g + \quadg)\cr}
$$
then following Fadeev-Popov we must calculate the determinant
$$
N = Vol(Weyl)~Det \lbrace -k(R_g + \quadg)\rbrace
$$

To do that we introduce the grasmaniann ghosts $\psi$ and $\bar\psi$
satisfying the  rules
$$
\lbrace\psi,\bar\psi\rbrace=0,\ \ \ \psi^2=\bar\psi^2=0
$$
and we can write $N$ in the form of path integral
$$
N = Vol(Weyl)~\int [D_g\psi][D_g\bar\psi] e^{{1 \over 2}k
\int d^2\xi \sqrt{g}\bar\psi(\xi) (R_g+\quadg)\psi(\xi)}\equiv
Vol(Weyl)~\tilde{N}
\eqn\due
$$

For what concerns the diffeomorphisms we follow the Distler-Kawai (DK)
procedure [2] and introduce a set of background
metrics parameterized by the moduli parameters,
$\hat g_{ab(\tau)}$. The gauge with respect to diffeomorphism's invariance
will be fixed parameterizing the metric by a Weyl scaling $\phi$
$$
g_{ab} = e^{\alpha\phi} \hat g_{ab(\tau)}
$$
and also the volume of dif\-fe\-o\-mor\-phism's group disappears
from the path integral \uno\ which now can be rewritten in the following way
$$
Z = \int [D_{\hat g}X^{\mu}][D_{\hat g}\phi]
  [D_{\hat g}b][D_{\hat g}c][d\tau]\tilde{N}
  e^{-S_M[X;\hat g]-S_{GH}[b,c;\hat g]-S_L[\phi;\hat g]}
$$
Here
$$
S_L[\phi;\hat g]={1\over 8\pi}\int d^2\xi\sqrt{\y}
(\hat g^{ab}\nabla_a\phi\nabla_b\phi - Q R_{\hat g}\phi)
$$
is the Liouville action without the term
$\mu_1\sqrt{\hat g}e^{\alpha\phi}$ already eliminated by setting the
appropriated value to the cosmological constant $\mu_0$.
Note also the scale parameter $\alpha$ which has been included
to have a standard kinetic term.

We can now match the two procedures paying attention to the fact that
having pa\-ra\-me\-te\-ri\-zed the metric as above,
we must effect the substitution
$$
g_{ab} = e^{\alpha\phi} \hat g_{ab}
\eqn\tre
$$
also into $\tilde{N}$ (from now on we will omit
the $\tau$ parameter).

Our ansatz, following DK, is that the measure for Weyl ghosts
transforms as the other measures,
that is with a jacobian determinant given by a Liouville action
$$
[D_g\psi][D_g\bar\psi] =
e^{S'_L[\phi;\hat g]}
[D_{\hat g}\psi][D_{\hat g}\bar\psi]
\eqn\quattro
$$
under a Weyl scaling of the metric $g_{ab} = e^{\alpha\phi} \hat g_{ab}$.
We can now substitute \tre\ in the determinant \due.

The relation between curvatures and between laplacians are
$$
\eqalign{
& R_g = e^{-\alpha\phi}(R_{\hat g}-\alpha\quadgg\phi)\cr
& \quadg = e^{-\alpha\phi}\quadgg\cr}
\eqn\cinque
$$
substituting these relations in the exponent in $\tilde{N}$
we have (remember that
$\sqrt g=e^{\alpha\phi}\sqrt{\hat g}$)
$$
\eqalign{
&{1 \over 2}k
\int d^2\xi \sqrt{g}\bar\psi(\xi) (R_g+\quadg)\psi(\xi) \cr
=&{1 \over 2}k
\int d^2\xi \sqrt{\hat g}e^{\alpha\phi}\bar\psi(\xi)[e^{-\a\phi}(R_{\hat g}-
\alpha\quadgg\phi)+e^{-\alpha\phi}\quadgg]\psi(\xi) \cr
=&{1 \over 2}k
\int d^2\xi \sqrt{\hat g}\bar\psi(\xi)(R_{\hat g}-
\alpha\quadgg\phi+\quadgg)\psi(\xi)\cr}
$$
so we can write $\tilde{N}$ in the following way
$$
\tilde{N} = \int [D_g\psi][D_g\bar\psi] e^{{1 \over 2}k
\int d^2\xi \sqrt{\hat g}\bar\psi(\xi)(R_{\hat g}-
\alpha\quadgg\phi+\quadgg)\psi(\xi)}
\eqn\sei
$$
where the relation \quattro\ is understood in eq.
\sei. As a consequence, we should
now transform the measure in eq. \sei\ inserting a Liouville determinant.
We shall have a unique Liouville action with undetermined parameters in
the partition function.

Thus we obtain the final path integral expression for $Z$
$$
Z = \int[D_{\hat g}X^{\mu}][D_{\hat g}\phi]
  [D_{\hat g}b[D_{\hat g}c][D_{\hat g}\psi][D_{\hat g}\bar\psi][d\tau]
  e^{-S_M[X;\hat g]-S_{GH}[b,c;\hat g]-S_L[\phi;\hat g]-
  S_W[\psi;\hat g]}
\eqn\sette
$$
where we have defined the Weyl action as
$$
S_W[\psi;\hat g] =
{1 \over 2}k
\int d^2\xi \sqrt{\hat g}\bar\psi\lbrace -(R_{\hat g}-
\alpha\quadgg\phi+\quadgg)\rbrace\psi
\eqn\otto
$$
We are interested in the total action for Liouville and Weyl ghosts
$$
\eqalign{
&S\equiv S_L[\phi;\hat g]+S_W[\psi;\hat g]\cr
=&{1\over 8\pi}\int d^2\xi\sqrt{\hat g}
(\hat g^{ab}\nabla_a\phi\nabla_b\phi - Q R_{\hat g}\phi)
+{k \over 2}
\int d^2\xi \sqrt{\hat g}\bar\psi\lbrace -(R_{\hat g}-
\alpha\quadgg\phi+\quadgg)\rbrace\psi\cr
=&{1\over 8\pi}\int d^2\xi\sqrt{\hat g}
(-\phi\qgg\phi -QR_{\hat g}\phi-q\bar\psi\qgg\psi
-q\bar\psi R_{\hat g}\psi+\alpha q\bar\psi\psi\qgg\phi)\cr}
$$
where we set
$$
\eqalign{
&q=4\pi k\cr
&\qgg=\quadgg\cr}
$$
that is
$$
S={1\over 8\pi}\int d^2\xi \sqrt{\hat g}
[-\phi\qgg\phi -QR_{\hat g}\phi-q\bar\psi\qgg\psi
-q\bar\psi R_{\hat g}\psi+\alpha q\bar\psi\psi\qgg\phi]
\eqn\nove
$$
Now the term in \nove\ relative to the Weyl ghosts
$(\bar{\psi},\psi)$ may be rewritten locally in the formal
way (we adopt a complex coordinate system)
$$
{q\over 8\pi}\sqrt{\hat g}[(\d\bar\psi)\dd\bar\psi-{1\over\d}(\d\bar\psi)\psi
R_{\hat g}+\alpha {1\over\d}(\d\bar\psi)\psi\d\dd\phi]
\eqn\dieci
$$
where $\d\dd=\qgg\vert_{loc}$ and ${1\over\d}$ denotes the resolvent
of the elliptic operator $\d$, i.e. formally
${1\over\d}(\d FIELD)\sim FIELD$.

Eq. \dieci\ has the traditional form of a ghost bc-system
after the identification
$$
\eqalign{
&b\equiv \d\bar\psi\cr
&c\equiv\psi\cr}
$$
Notice that the conformal dimension of $\d\bar\psi$ and $\psi$
are 1 and 0 respectively. We can then bosonize the action \dieci\
following [5] by introducing a free scalar field $\varphi$ and
expressing the ghost fields $(\bar{\psi},\psi)$ as
$$
\eqalign{
&b=\partial exp(i\v)\cr
&c=exp(-i\v)\cr}
\eqn\undici
$$
In this way we found
$$
\eqalign{
&{q\over 8\pi}\sqrt{\hat g} [(\d\v\dd\v-i\v R_{\hat g})-iA{1\over\d}(\d\v)
R_{\y}+i\alpha A{1\over\d}(\d\v)\d\dd\phi]\cr
=&{q\over 8\pi}\sqrt{\hat g} [-\v\d\dd\v-i(1+A)\v R_{\hat g}+
2i{\alpha A\over 2}\v\d\dd\phi ]\cr}
\eqn\dodici
$$
where we used
$$
bc=iA(\d\v)
\eqn\tredici
$$
Here $A$ is a renormalization factor which shall play the role of a new
bare parameter.

If now we define the free unrenormalized parameter
$$
\eqalign{
&(1+A)\equiv \tilde Q\cr
&{\alpha A\over 2}\equiv B\cr
&q\equiv 8\pi\cr},
\eqn\quattordici
$$
where $\alpha$ is understood as a gauge (bare) parameter
not to be confused with the Liouville field's moment,
the equation \dieci\ in a general frame becomes
$$
\sqrt{\y}[-\v\qgg\v+2iB\v\qgg\phi-i\tilde Q\v R_{\y}]
$$
After the Wick rotation of field $\v$, $\v\to i\v$, in order to get
a Liouville field, we obtain
$$
\sqrt{\y}[\v\qgg\v-2B\v\qgg\phi+\tilde Q\v R_{\y}]
\eqn\quindici
$$
Putting eq. \quindici\ in \nove\ we have the final form
for the Liouville-Weyl action in matrix notation
$$
S = \int d^2\xi \sqrt{\y} [-M_{i,j}\Phi^i\qgg\Phi^j-Q_i\Phi^i R_{\y}]
$$
$$
\eqalign{
&\Phi^i\equiv(\Phi^1,\Phi^2)=(\phi,\v)\cr
&Q_i\equiv(Q_1,Q_2)=(Q,-\tilde Q)\cr
&M_{i,j}\equiv\left(\matrix{1&B\cr
B&-1\cr}\right)\cr}
\eqn\sedici
$$

In the following we shall regard the three parameters
$Q_1,Q_2$ and $B$ as renormalized parameters to be fixed
by conformal invariance and by the requirement to get the
DK-regime in the limit $B\to 0$.

The holomorphic energy-momentum tensor associated with
\sedici\ reads locally as
$$
T = -{1\over 2}[M_{i,j}\d\Phi^i\d\Phi^j+Q_i(\d^2)\Phi^i]
\eqn\diciassette
$$
The correspondent central charge is given by
$$
c_{\scriptscriptstyle{W+L}}=2+3M^{i,j}Q_iQ_j
\eqn\diciotto
$$
where $M^{i,j}$ is the inverse matrix of $M_{i,j}$.

Our key idea is that the model \sedici\ replaces the DK Liouville
action when one consider the coupling 2D-QG to conformal matter fields
with central charge $c_m$, for any value of $c_m$.
To show that, we calculate string susceptibility of the theory and we
verify that its value is real for all $c_m$.

In particular let us consider the coupling of \sedici\
with $D$ free massless scalar matter fields, so that before the coupling
their central charge is $c_m=D$.
We start by noting that, as in DK, the gravitational
coupling implies that the matter vertex
operator $V_{(a)}$ had a "gravitational dressing" given by
$$
V_{(a)}\to V_{(a)}e^{\alpha_1\Phi_1}\equiv
V_{(a)}e^{\alpha\phi},
\eqn\diciannove
$$
so that, if $V_{(a)}$ is the identity operator,
conformal invariance requires that $e^{\alpha\phi}$ is a (1,1)
conformal field and its conformal dimension as computed by
\diciassette\
$$
\Delta[e^{\alpha_1\phi}]=-{1\over 2} \alpha_1 M^{1,j}(\alpha_j+Q_j),
\eqn\venti
$$
must be equal to 1. This fact will be used to compute
parameter $\a_1$.

To calculate string susceptibility we use the expression [2]
$$
\Gamma = \chi(h) {Q_1\over 2 \alpha_1} + 2
\eqn\susc
$$
where $\chi(h)=2(1-h)$ is the Euler characteristic of the surface,
hence we must determinate the values of parameters $Q_1$ and $\a_1$.
We begin by noting that, as well known,
Weyl invariance of the whole system requires that
$$
c_{\scriptscriptstyle{W+L}}+D-26=0
\eqn\ventuno
$$
Using \diciotto\ in eq. \ventuno\ we find
$$
Q_1^2+2BQ_1Q_2-Q^2_2+{1+B^2\over 3}(D-24)=0
\eqn\ventidue
$$
To solve this equation we adopt the following parametrization
$$
\eqalign{
&Q_2=\sqrt{{1\over 3}+\lambda (B,D)}\cr
&\lim_{B\to 0}\lambda (B,D)=0\cr}
\eqn\ventitre
$$
In this parametrization $Q_1$ is given by (see appendix B)
$$
Q_1=-{1\over \sqrt{3}}\bigl[ B\sqrt{1+3\lambda (B,D)}-
\sqrt{(1+B^2)(3\lambda(B,D)+25-D)}\bigr]
\eqn\ventiquattro
$$
where the sign is chosen to have the standard semiclassical
limit for $D\to -\infty$ (see appendix C).

We can calculate $\alpha_1$ using \venti\
(remember that $\alpha_2=0$).
$$
-{1\over 2} \alpha_1 M^{1,j}(\alpha_j+Q_j)=1
\eqn\venticinque
$$
and we get (see appendix B)
$$
\alpha_1=-{1\over 2\sqrt{3}}
\biggl[\sqrt{(1+B^2)(25-D+3\lambda(D,B))}
-\sqrt{(1+B^2)(1-D+3\lambda(D,B))}\biggr]
\eqn\ventisei
$$
where the sign have been chosen again to have the semiclassical
limit for $D\to -\infty$ (see appendix C).

String susceptibility $\Gamma$ can now be calculated
by using the expression \susc\ and known values
for $Q_1$ and $\alpha_1$
in $\lambda$ parametrization. We find
$$
\eqalign {
\Gamma = {1-h\over 12(1+B^2)}&\biggr\{ B \sqrt{1+B^2}
\Bigl[\sqrt{(1+3\j)(25-D+3\j)}+\sqrt{(1+3\j)(1-D+3\j)}\Bigr]+\cr
&-(1+B^2)\Bigl[25-D+3\j+\sqrt{(25-D+3\j)(1-D+3\j)}\Bigr]\biggr\}\cr}
\eqn\ventisette
$$
We fix a particular parametrization choosing the following form for the
function $\j(D,B)$
$$
\j(D,B) = {B\over 3} D,
\eqn\ventotto
$$
in fact we can see that in this parametrization the reality of
$\Gamma$ is ensured for all $D>0$ simply requiring that $B>1$,
as shown in appendix D, and the DK-limit is reached for $B\to 0$.

Although we have no informations on B at this level, we see that
assuming that DK regime is achieved from eq. \ventisette\ just requiring
the condition $D\le 1$, then $B$ must be a step function of $D$ as
$B\sim\overline{B}\theta (D-1)$, where $\theta(x)=0$ for $x\le 0$ and
$\overline B$ is a generic real constant greater than one.

\endpage

\title{APPENDIX A}

We show that for every metric $g_{ab}$ there exists just one metric
$\y_{ab}$ so that
$$
\g=e^\sigma\h ~~and~~R_{\y}=-1
$$
To demonstrate existence we start by writing the relation between
scalar curvatures of the metrics $\g$ and $\h$
$$
R_g=e^{-\sigma}(R_{\y}-\y^{ab}\hat\nabla_a\partial_b\sigma)
\eqno {(A1)}
$$
and vice versa
$$
R_{\y}=e^{\sigma}(R_g+g^{ab}\nabla_a\partial_b\sigma)
\eqno {(A2)}
$$
Now by requiring that $R_{\y}=-1$ we get the equation [6]
$$
R_g(x,y)+g^{ab}(x,y)\nabla_a\partial_b\sigma(x,y)=-e^{-\sigma(x,y)}
$$
We can reduce complexity by considering that, at least locally,
$\g$ is conformally Euclidean, that is $\g=e^\rho\delta_{ab}$.
{}From ($A$2) we obtain locally
$$
-1=R_{\y}=e^{\sigma'}(R_{\delta}+\delta^{ab}\partial_a\partial_b\sigma')
$$
where $\sigma'=\rho\sigma$, so, at least locally, we must deal with
the solvable equation
$$
(\partial_x^2+\partial_y^2)\sigma'(x,y)=-e^{-\sigma'(x,y)}.
$$

To verify unicity we proceed via the reductio ab absurdo,
allowing the existence
of two metrics $\g'$ and $\g''$ both conformal equivalents to
$\g$ and having scalar curvature equal to $-1$, that is
$$
\eqalign{
&\g=e^{\sigma'}\g'~~and~~R_{g'}=-1\cr
&\g=e^{\sigma''}\g''~~and~~R_{g''}=-1\cr}
$$
so we can write
$$
\g''=e^{\sigma'-\sigma''}\g'
$$
Now using ($A$1) we can write the following relations
$$
\eqalign{
&\g'=e^{-\sigma'}\g\cr
&R_{g'}=e^{\sigma'}(R_g+g\nabla_a\partial_b\sigma')=-1\cr}
\eqno {(A3)}
$$
$$
\eqalign{
&\g''=e^{-\sigma''}\g\cr
&R_{g''}=e^{\sigma''}(R_g+g\nabla_a\partial_b\sigma'')=-1\cr}
\eqno {(A4)}
$$
$$
\eqalign{
&\g''=e^{\sigma'-\sigma''}\g'\cr
&R_{g''}=e^{-(\sigma'-\sigma'')}
(R_g'-g'^{ab}\nabla'_a\partial'_b(\sigma'-\sigma''))=-1\cr}
\eqno {(A5)}
$$
from the first two relations we obtain
$$
\eqalign{
&R_g+\nabla_a\d_b\sigma'=-e^{-\sigma'}\cr
&R_g+\nabla_a\d_b\sigma''=-e^{\sigma''}\cr}
$$
so
$$
\nabla_a\d_b (\sigma'-\sigma'')=e^{-\sigma''}-e^{-\sigma'}
\eqno {(A6)}
$$
To use the last relations we notice that
$$
\nabla_a\d_b\sigma=g'_{ab}\nabla'_a\partial'_b\sigma
$$
If the dimension of the world sheet is 2
and $\sigma$ is a conformal tensor of weight
s=0, the latter should transform as
$$
\sigma'=\Omega^s\sigma
$$
under a coordinates transformation $z'^a=z'^a(z)$ for which
$\g'=\Omega\g$ [7].
This conditions hold in present case because we are dealing with a
transformation of the metric alone, so that $z'^a=z^a$ and being
$\sigma$ a scalar field $\sigma'(z')=\sigma(z)$
so that $\sigma'=\sigma$.

{}From ($A$5) we can now write
$$
\nabla_a\d_b (\sigma'-\sigma'')=e^{(\sigma'-\sigma'')}-1
$$
which compared with ($A$6) gives
$$
e^{-\sigma''}-e^{-\sigma'}=e^{(\sigma'-\sigma'')}-1
$$
we can solve this equation by setting
$$
\eqalign{
&x=e^{-\sigma'}\cr
&y=e^{-\sigma''}\cr}
$$
so we obtain
$$
y-x={y-x\over x}
$$
which solutions are x=y or x=1, that is
$$
\sigma'=\sigma''
$$
or
$$
\sigma'=0
$$
which both confirm absurdity of initial assumption and complete
the demonstration of unicity.

\endpage

\title{APPENDIX B}

To proceed to the determination of parameters $Q_1$, $Q_2$ and $\alpha_1$
we must first determine the inverse matrix $M^{i,j}$ of
$M_{i,j}$ defined in \sedici\
$$
M^{i,j}={1\over 1+B^2}\left(\matrix{1&B\cr
B&-1\cr}\right)
\eqno {(B1)}
$$
Then we can develop equations \ventidue\ and \venticinque\ obtaining
$$
\eqalign{
&Q_1=-BQ_2\pm {1\over\sqrt{3}}\sqrt{(1+B^2)(3Q_2^2+24-D)}\cr
&\a_1^2+(Q_1+BQ_2)\a_1+2(1+B^2)=0\cr
&\a_1=-{1\over 2\sqrt{3}}\biggr[\pm\sqrt{(1+B^2)
(3Q_2^2+24-D)}\mp\sqrt{(1+B^2)(3Q_2^2-D)}\bigg]\cr}
$$
Now inserting the parametrization \ventitre\ we get
equations \ventiquattro\ and \ventisei\ .
\endpage

APPENDIX C

We verify that the parametrization \ventitre\ is compatible with
DK-regime. Specifically we get DK-regime in the limit for
$B\to 0$.

For $B\to 0$ we have
$$
\eqalign{
&\j\to 0\cr
&Q_2={1\over 3}\cr
&Q_1\to \sqrt{{25-D\over 3}}\cr}
$$
which is exactly DK values for the central charge.
Moreover we get the following values for $\a_1$ and $\Gamma$
$$
\eqalign{
&\alpha_1=-{1\over 2\sqrt{3}}
\Bigl[\sqrt{25-D}-\sqrt{1-D}\Bigr]\cr
&\Gamma={(1-h)\over 12}\Bigl[ D-25-\sqrt{(25-D)(1-D)}\Bigr]\cr}
$$
which are again DK values.
\endpage

\title{APPENDIX D}

We want to verify the reality of $\Gamma$ as given by \ventisette\ ,
when $\j$ is fixed by the parametrization \ventotto\ .
Reality of $\Gamma$ is obtained by simultaneously satisfying
$$
\eqalign{
&(1+DB)(25-(1-B))\geq 0\cr
&(1+DB)(1-(1-B))\geq 0\cr
&(1-(1-B))(25-(1-B))\geq 0\cr}
$$
requiring that $B>1$ and introducing the parameter
$$
\eqalign{
&a=B\cr
&b=B-1\cr}
$$
we can solve the system
$$
\eqalign{
&(1+aD)(25+bD)\geq 0\cr
&(1+aD)(1+bD)\geq 0\cr
&(1+bD)(25+bD)\geq 0\cr}
\eqno {(D1)}
$$
where the following relations between the parameters hold
$$
a>b>0
$$
$$
-{25\over b} < -{1\over b} < -{1\over a} < 0
$$
We can so write the solutions of the equations in system ($D$1)
$$
\eqalign {
&D\leq -{25\over b}~;~D\geq -{1\over a}\cr
&D\leq -{1\over b}~;~D\geq -{1\over a}\cr
&D\leq -{25\over b}~;~D\geq -{1\over b}\cr}
$$
and the solution of the system is
$$
D\leq -{25\over b}~;~D\geq -{1\over a}
$$
we are interested in positive solutions, that is for
$D\geq -{1\over a}$, which traduces in
$$
D\geq -{1\over B}
$$
so being $B>1$ we have verified that system ($D$1) is
satisfied for all $D>0$.

\endpage

ACKNOWLEDGEMENT

K. Yoshida would like to acknowledge illuminating discussions with
K. Fujikawa and H. Kawai.

REFERENCES

\parindent=0.pt

[1] A. M. Polyakov, Phys. Lett. B103 (1981) 207

[2] J. Distler and H. Kawai, Nucl. Phys. B321 (1989) 509

[3] O. Alvarez, Lectures at the Workshop on Unified String Theory held
in the Institute of Theoretical Physics at University of California,
Santa Barbara, August 1985

[4] E. D'Hoker and D. H. Phong, Nucl. Phys. B269 (1986) 205

[5] T. Eguchi and H. Ooguri, Phys. Lett. B187 (1987) 127

[6] Itzykson and Drouffe, "Theorie statistique des champs", Cambridge
Univ. Press (1990) 361

[7] R. M. Wald, "General Relativity", The University of Chicago Press
(1984) 446

\end